\documentclass[12pt,preprint]{aastex}
\usepackage{graphicx}
\usepackage{ulem}
\usepackage{rotating}
\usepackage{lscape}
\usepackage{natbib}
\begin{document}

\title{Time-monitoring Observations of the Ro-Vibrational Overtone CO
  bands in Young Stars}

\author{J. A. Eisner, G. H. Rieke, M. J. Rieke, K. M. Flaherty, T. J. Arnold,
  J. M. Stone, S. R. Cortes, E. Cox}
\affil{Steward Observatory, The University of Arizona, 933 N. Cherry
  Ave, Tucson, AZ 85721}
\email{jeisner@email.arizona.edu}

\and

\author{C. Hawkins, A. Cole, S. Zajac, A. L. Rudolph}
\affil{Department of Physics and Astronomy, California State
  Polytechnic University, 3801 W Temple Ave, Pomona, CA 91768}


\begin{abstract}
We present near-IR spectra of a sample of T Tauri, Herbig Ae/Be, and
FU Ori objects.   Using the FSPEC instrument on the Bok 90-inch telescope, we
obtained $K$-band spectra with a resolution of $\approx 3500$.  
Here we present spectra of the $\Delta v=2$--0 and
$\Delta v=3$--1 bandheads of ro-vibrational transitions of carbon
monoxide.  We observed these spectra over multiple epochs spaced by a
few days and approximately one month.  Several of our targets show CO
emission or absorption features.  However we see little evidence of
variability in these features across multiple epochs.  We compare our
results with previous observations, and discuss the physical
implications of non-variable CO emission across the sampled timescales.
\end{abstract}

\section{INTRODUCTION \label{sec:intro}}
Young stellar objects have long been known as
variable from UV to IR wavelengths 
\citep[e.g.,][]{JOY42,HERBST+94,SKRUTSKIE+96}.  Several different mechanisms
have been proposed to explain various types of monochromatic and color
variability: 1) variable stellar flux, for example due to magnetic cool
spots or accretion hot spots; 2) variable extinction, perhaps caused by
gas and dust orbiting in a circumstellar disk; and 3) variations
intrinsic to the inner circumstellar disk, such as changing inner radii
or temperatures.  


Multi-wavelength photometric variability in a few objects has
been used to constrain these scenarios
\citep[e.g.,][]{CHS01,EIROA+02,FORBICH+07,FLAHERTY+12}.    
Optical spectroscopic
monitoring has also provided constraints on variable stellar or
accretion properties \citep[e.g.,][]{EISNER+10b,MENDIGUTIA+11}.  Spectroscopic
monitoring in the near-IR---the peak spectral region for thermal
emission in the inner disk---can probe disk variability in particular.

The ro-vibrational overtone transitions of CO can be used to probe
changes in the structure of hot, optically thin disk regions.
Overtone transitions of CO have been detected in emission toward a
number of young stars
\citep{CARR89,NAJITA+96,NAJITA+00,NAJITA+06,NAJITA+09,
BISCAYA+97,THI+05,BRITTAIN+07,BERTHOUD+07,BERTHOUD08}. 
Because these transitions 
require high excitation temperatures, we expect this emission to trace
disk regions close to the star (within $\sim 0.1$ AU).  Analyses of
spectrally resolved line profiles confirm that CO overtone
emission generally arises in the inner regions of Keplerian disks
\citep[e.g.,][]{NAJITA+96,NAJITA+09,THI+05}.
 
CO overtone emission originates from 
stellocentric radii in between star/accretion flow interfaces
and dust sublimation radii.  While accretion rates and dust
continuum emission are both known to vary for young stars
\citep[e.g.,][]{CHS01,EISNER+10b}, less is known about the variability
of inner disk gas.  CO emission can therefore provide a
valuable diagnostic of inner disk variability.
Variability in the emission has been observed in some objects,
with emission features disappearing entirely in some cases
\citep[e.g.,][]{BISCAYA+97,NAJITA+00}.

CO absorption is also observed toward some young stars.  The CO
overtone bandheads are seen in absorption from
the photospheres of late-type stars.  Since these absorption
features occur in stellar photospheres, one might expect limited
variability.  However if this absorption is superposed on emission
from circumstellar disk gas, then variability might be observed.

Absorption features may also arise purely from circumstellar disk gas
in special
cases.  Emission from hot circumstellar matter may be
absorbed by cooler CO along the line-of-sight.  For example, a
circumstellar disk midplane heated to a temperature of several
thousand K (e.g., by viscous dissipation in accreting matter) can
produce emission that is then absorbed by a cooler disk atmosphere.
Such models can produce CO absorption spectra
much like those produced by cool stellar photospheres
\citep{CALVET+91}.  

This
scenario is thought to apply to the FU Orionis sources, a class of
young stars surrounded by actively accreting, very hot disks
\citep[e.g.,][]{HK85}.  Indeed, the narrow line widths and spatial
distributions of CO emission in FU Ori stars (including the two in
the sample presented here) confirm that the CO absorption features
arise in extended circumstellar disks and not in stellar photospheres
\citep{HHC04,GAR08,ZHU+09,EH11}. 
The CO absorption features around these
objects---and their potential variability---can therefore probe inner
disk conditions.

Here we present time-monitoring observations of the CO overtone
emission in a number of young stars.  Our sample includes FU Ori
sources (V1515 Cyg and V1057 Cyg), T Tauri stars (AS 205 N, V 1002
Sco, V2508 Oph, AS 209), and Herbig Ae/Be objects (MWC 863, 51 Oph,
MWC 275, V1685 Cyg, AS 442, and V1331 Cyg).  In the next section
we describe the observing and data reduction procedure.  Results and
discussion of
the multi-epoch observations are described in Sections
\ref{sec:results} and \ref{sec:disc}.

\section{OBSERVATIONS AND DATA REDUCTION \label{sec:obs}}
We used the near-IR spectrograph FSPEC at the Bok 90-inch telescope to
monitor the CO emission around a sample of young stars (Table
\ref{tab:sample}).  We used a grating with 600 grooves per mm,
providing a resolving power of
$\lambda/\Delta \lambda = 7400$ (per pixel).
FSPEC provides a $2.4'' \times 96''$ slit, with $1.2''$ pixels.  The
slit is $\sim 2$ pixels wide, and the actual spectral resolution is
closer to 3500.

In our observations we nodded along the slit, observing each object
at 5--6 distinct slit positions.  Total integration times (including
all slit pointings) for each observation are listed in Table \ref{tab:sample}.
Observations of target objects were interleaved with observations of
telluric calibrators.  We selected calibrators with spectral types
earlier than A0.  Because these calibrators do not show CO absorption
(or emission)
features in their photospheric spectra, no spurious signals are
introduced into the target spectra.

We developed an IDL-based data reduction pipeline to produce
calibrated spectra.  The pipeline works on an entire night of data as
a block, producing telluric-corrected, wavelength-calibrated
spectra for each object observed during the night.  Because we work
with the entire night of data at once, we can use a weighted sum of
all the data to produce higher signal-to-noise in certain calibrations.  

The first step in the data reduction procedure is the production of
median flat and dark images, and a bad pixel mask.  The bad pixel mask
specifies the location of hot, dead, and ``flaky'' pixels, where flaky
pixels are those showing an abnormally high standard deviation between frames.
We correct bad pixels in the raw data by replacing values flagged in
the bad pixel mask with the median values of surrounding pixels.

After bad-pixel correction we perform sky
subtraction for every observed object, including targets and calibrators.
For each position of an object within the slit, data from adjacent
slit positions are subtracted.  Discarding values less than zero
yields a sky-subtracted image of only the slit position of interest.  
We then divide by the flat.

At this point, we calculate a ``sky frame''.  We compute this frame
from a weighted sum of all observations for the entire night.  Since
we are interested in sky emission, the weighting is higher for
observations with longer integration times.  Because every observation
contains sky emission lines at the same wavelengths, these add
coherently.  Spectra from observed objects, which are nodded along the
slits, do not add coherently.  The result is an image of the sky
emission lines, relatively ``uncontaminated'' by spectra of observed
targets and calibrators.

We use the sky frame to determine the slope of sky emission lines
across the detector.  We need this information in order to determine
how a spatial shift along the slit affects the wavelength direction.
We determine the spatial positions of spectra at different slit
positions by fitting Gaussians to spectra projected into the spatial
dimension, then shift them along the determined slope to 
create a single, summed 2D spectrum.  To ensure that
sources and calibrators share a common wavelength calibration, we
always shift the spectra to the same spatial position, typically in
the middle of the slit.

From the summed, 2D spectrum,
we determine the spectral trace, which is not exactly rectilinear on
the detector, by fitting Gaussians to slices through each spatial
position.  We extract spectra from $\pm 2.5$ pixels around the trace
positions.  After extraction we are left with a 1D spectrum for each
observed object.

Telluric calibration is accomplished by dividing the 1D spectrum for a
target by the 1D spectrum of its corresponding calibrator.  We
multiply this result by a
Kurucz model atmosphere appropriate for the calibrator's spectral type
to remove overall slopes associated with the calibrator spectrum.
When more than one calibrator is observed, or if one is observed
multiple times, we determine an average, calibrated spectrum.

Finally, we perform a wavelength calibration.  We determine a solution
for an entire night of data and then apply it to every observed
object.  The wavelength solution is computed for the sky frame
(discussed above).  Using the same trace
determined for the extracted target and calibrator spectra, we
extract a 1D sky spectrum.  We identify telluric emission lines, of known
wavelengths, in the sky spectrum, and compute a polynomial wavelength
solution.  We typically use a third-order polynomial, which provides
solutions with slightly lower residuals than a second-order polynomial fit.

One might wonder why we compute a wavelength solution for the entire
night rather than for individual objects.  If the wavelength solution
varies throughout the night, then solutions for individual objects
would be preferred.  However if the wavelength solution is stable,
then low signal-to-noise in observed sky lines for some observations
might limit the accuracy of the wavelength calibration.  

We determine the stability of the wavelength solution by measuring the
wavelengths of telluric lines across multiple sources and epochs.
Using the average sky frame for wavelength calibration produces a
smaller scatter in the source-to-source and epoch-to-epoch wavelength
determinations of telluric lines than we see if we compute wavelength
solutions for individual objects.  The stability of FSPEC over a night
is not surprising given the exceptional stability and negligible
flexure of the instrument.

Since our observations were not taken under photometric conditions, we
do not attempt to associate real continuum fluxes with our reduced
spectra.  Rather, we median-subtract each spectrum so that all spectra
for a given object have the same continuum level.  We thus ignore any
potential variability in the continuum flux of our targets.  Such
variations would not be surprising given the known infrared
variability of many young stars \citep[e.g.,][]{SKRUTSKIE+96,EIROA+02}.

\section{RESULTS \label{sec:results}}
\begin{figure}
\plotone{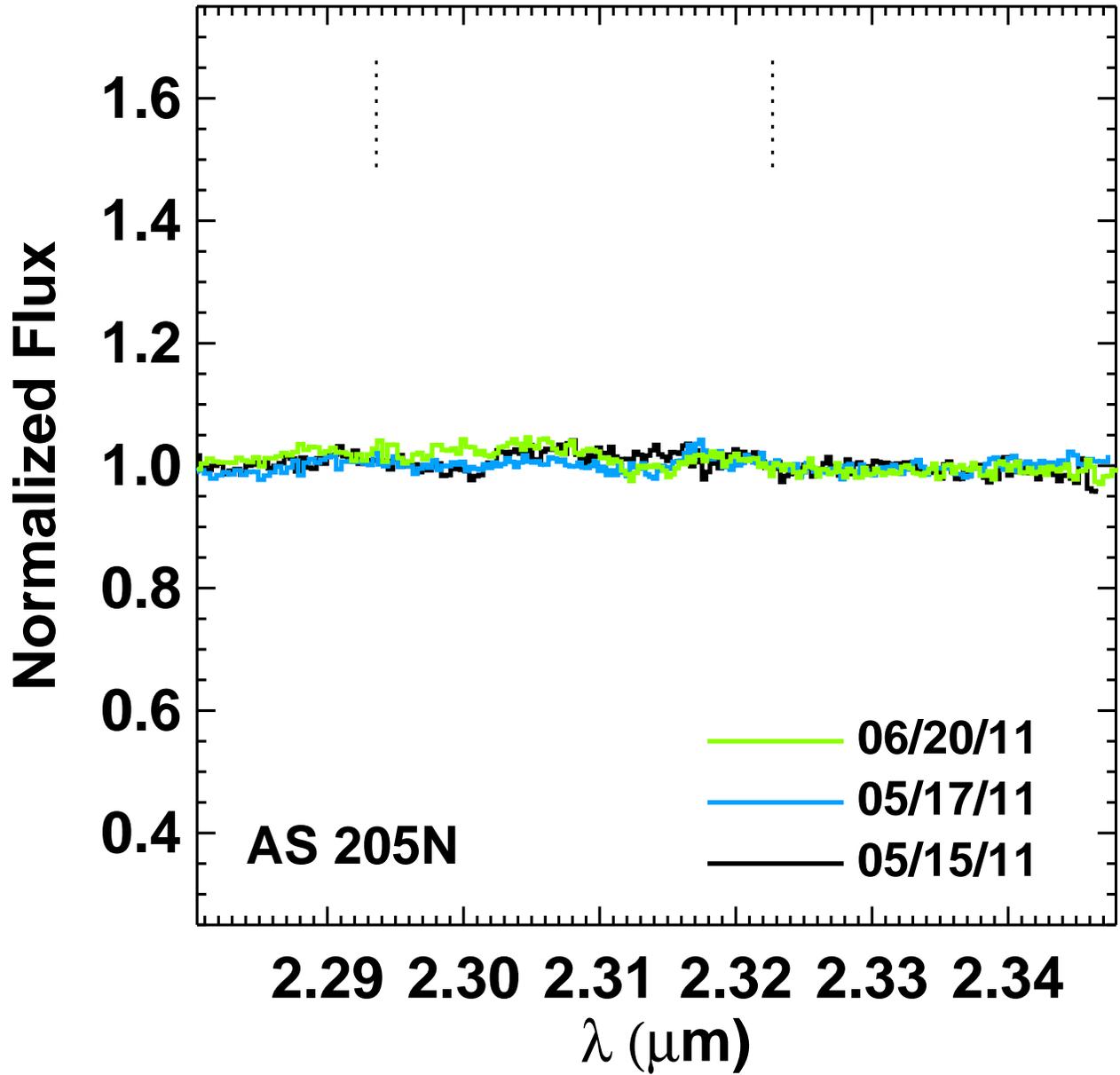}
\caption{Spectra of AS 205 N.  The wavelengths of the $v = 2
  \rightarrow 0$ and $v = 3\rightarrow 1$ overtone bandheads of CO are
  indicated by dashed vertical lines.
\label{fig:as205}}
\end{figure}

\begin{figure}
\plotone{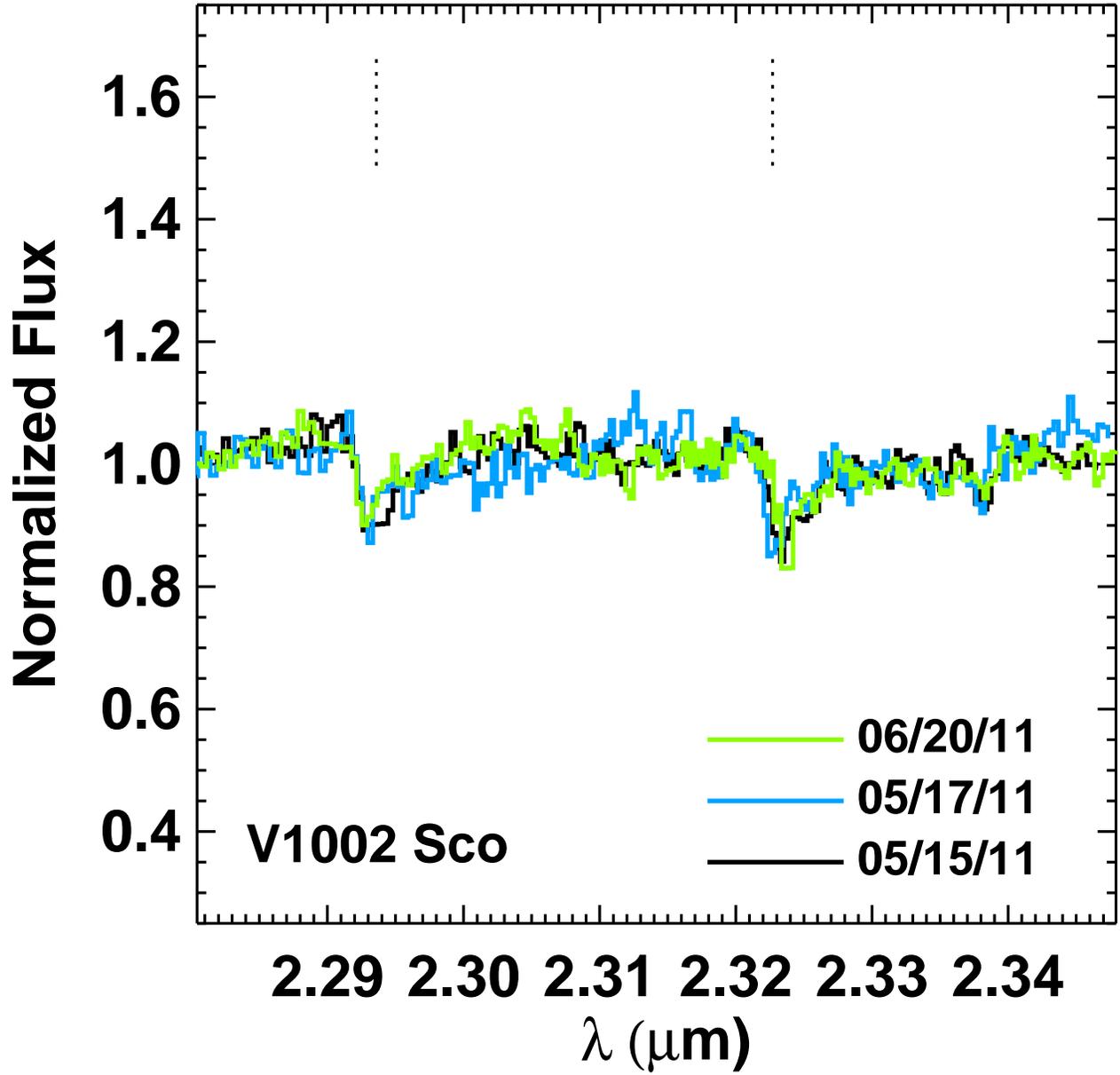}
\caption{Spectra of V1002 Sco.
\label{fig:v1002}}
\end{figure}

\begin{figure}
\plotone{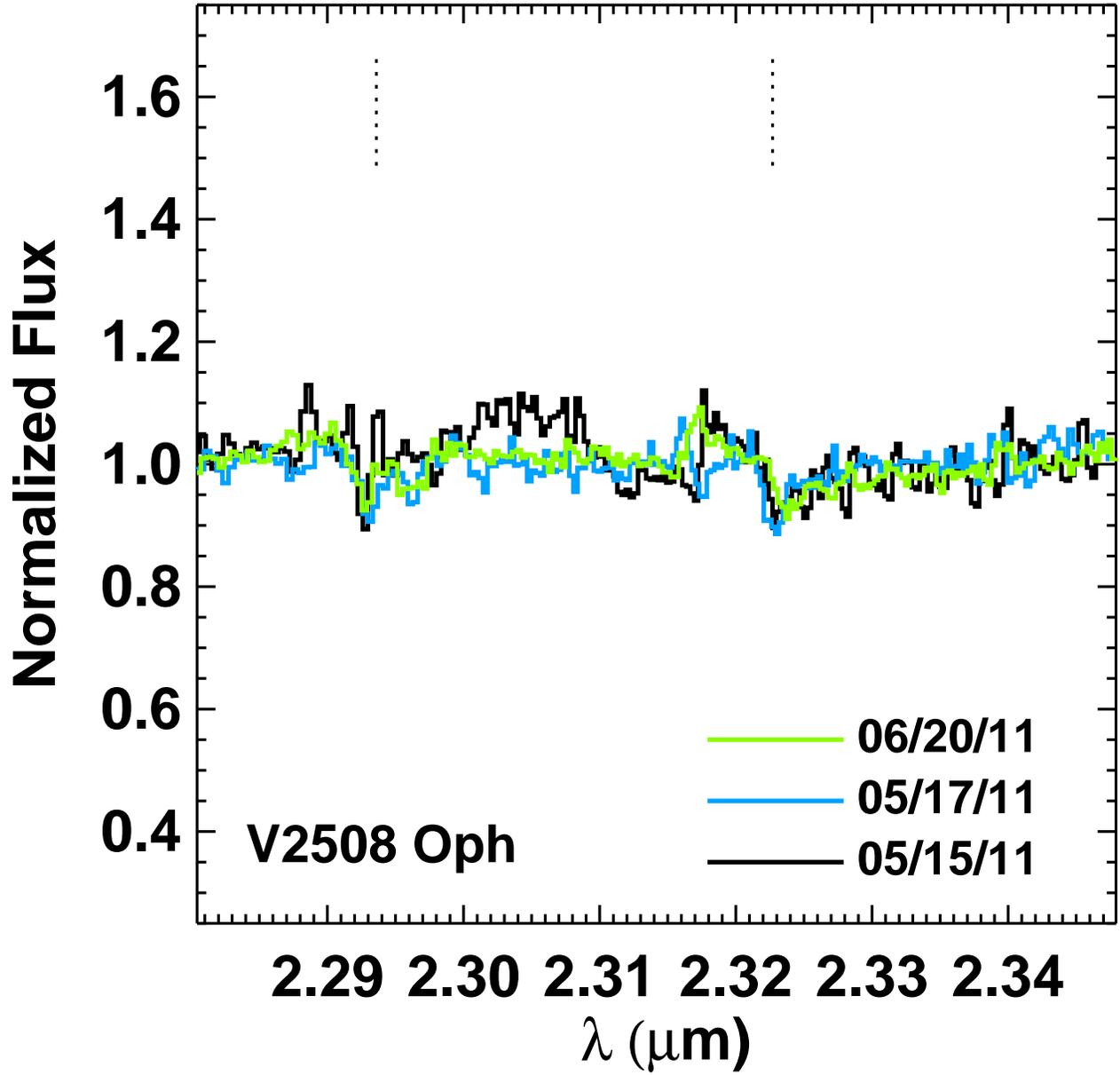}
\caption{Spectra of V2508 Oph.
\label{fig:v12508}}
\end{figure}

\begin{figure}
\plotone{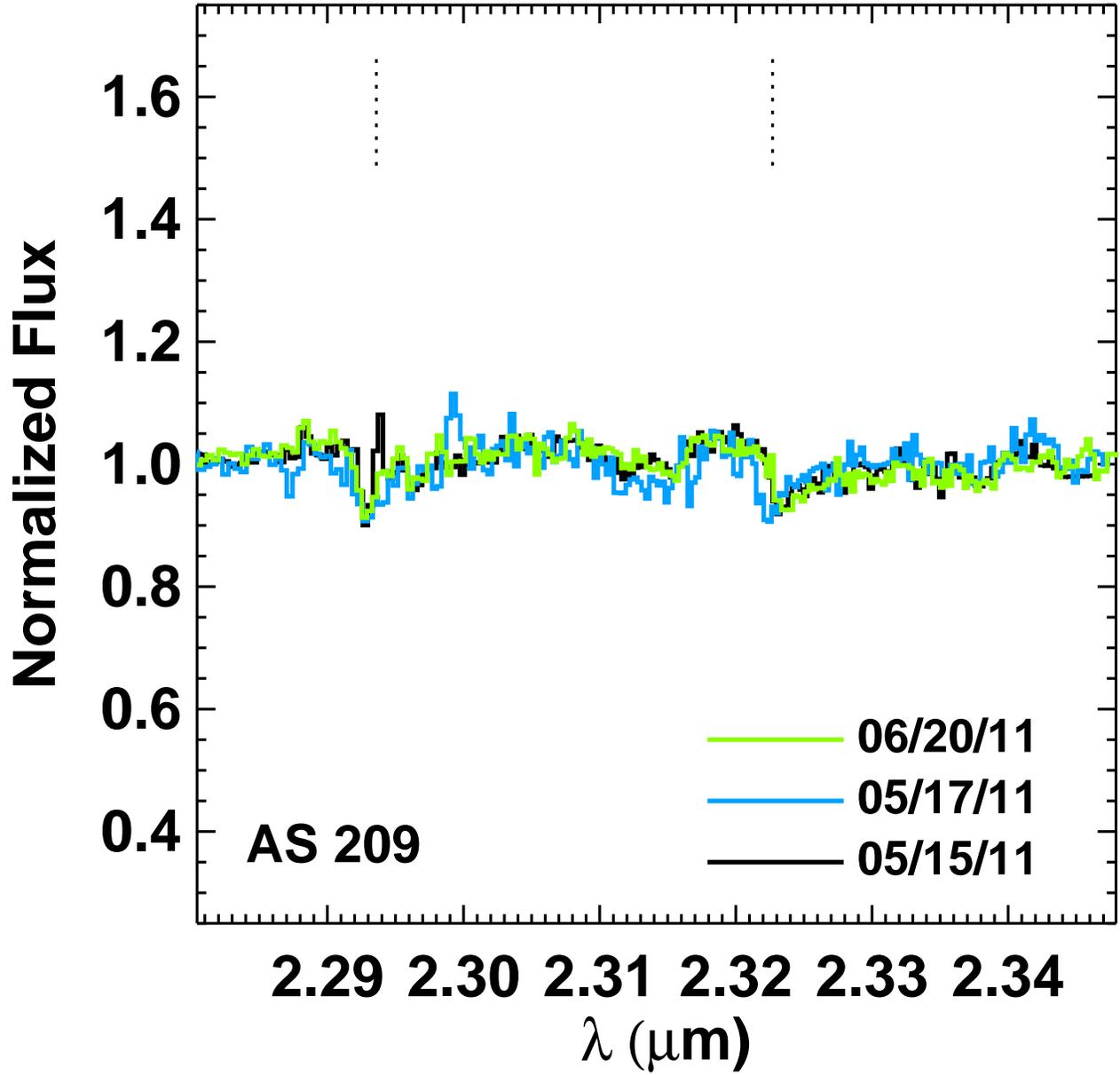}
\caption{Spectra of AS 209.
\label{fig:as209}}
\end{figure}

\begin{figure}
\plotone{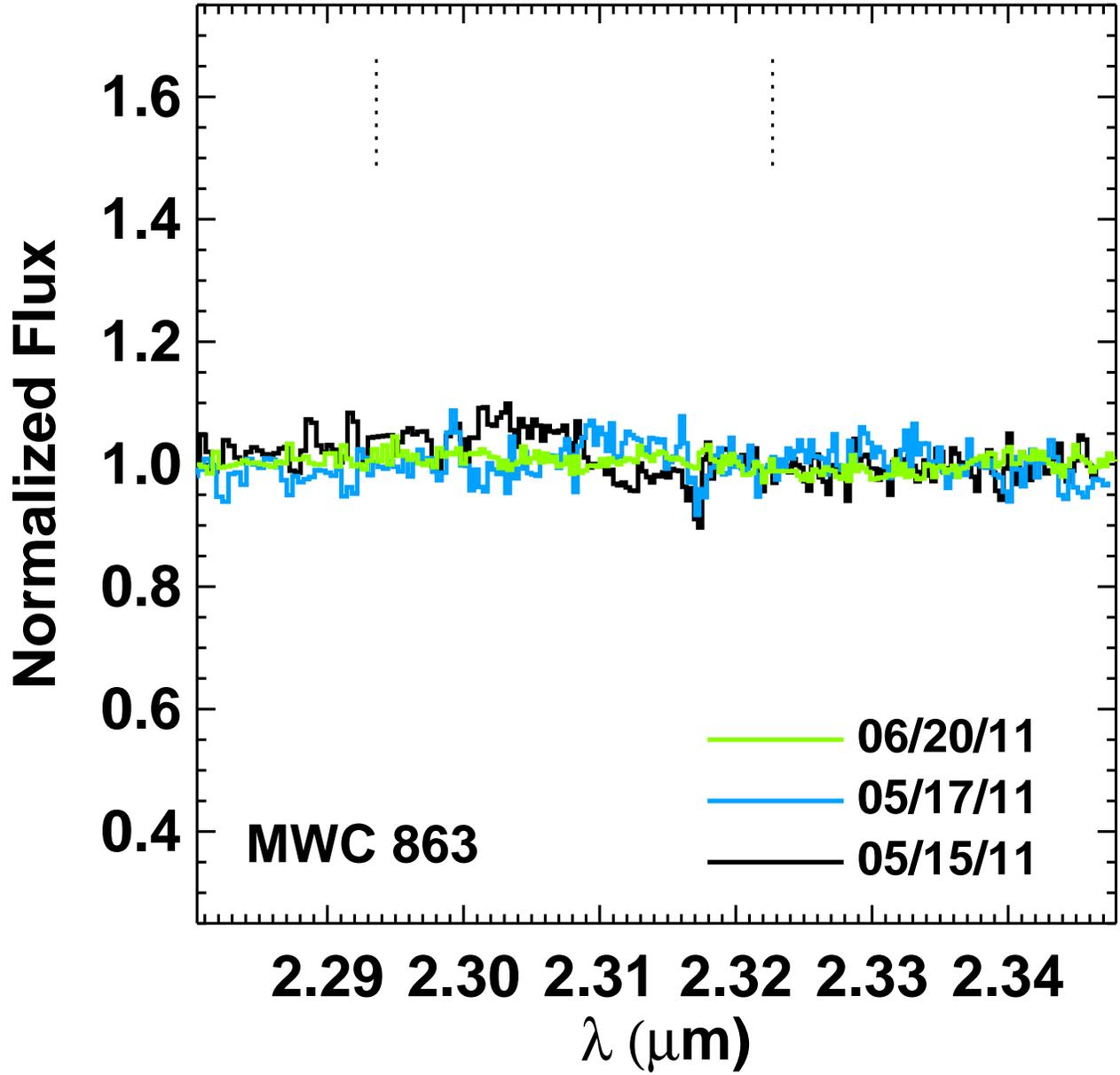}
\caption{Spectra of MWC 863.
\label{fig:mwc863}}
\end{figure}

\begin{figure}
\plotone{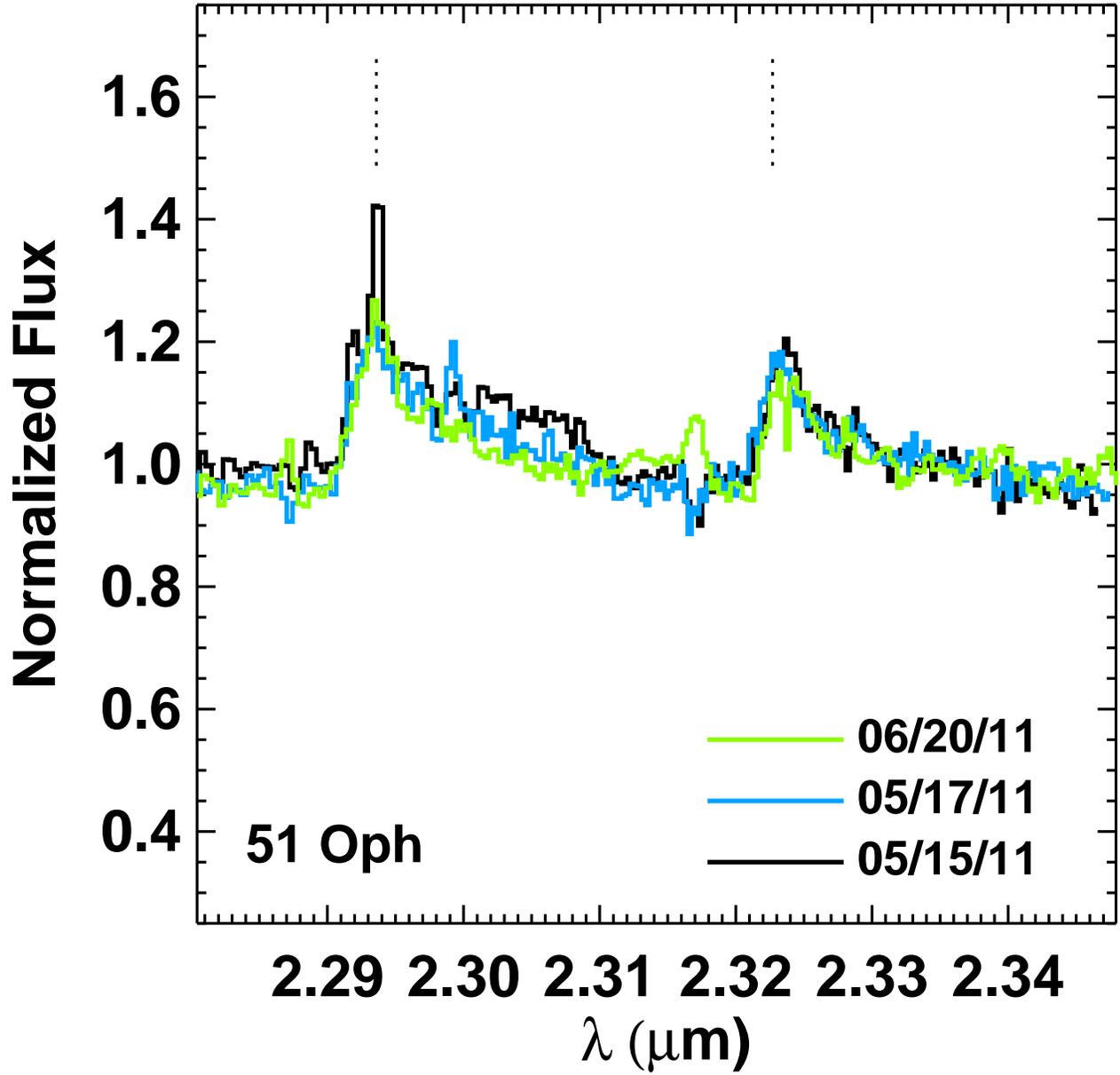}
\caption{Spectra of 51 Oph.
\label{fig:51oph}}
\end{figure}

\begin{figure}
\plotone{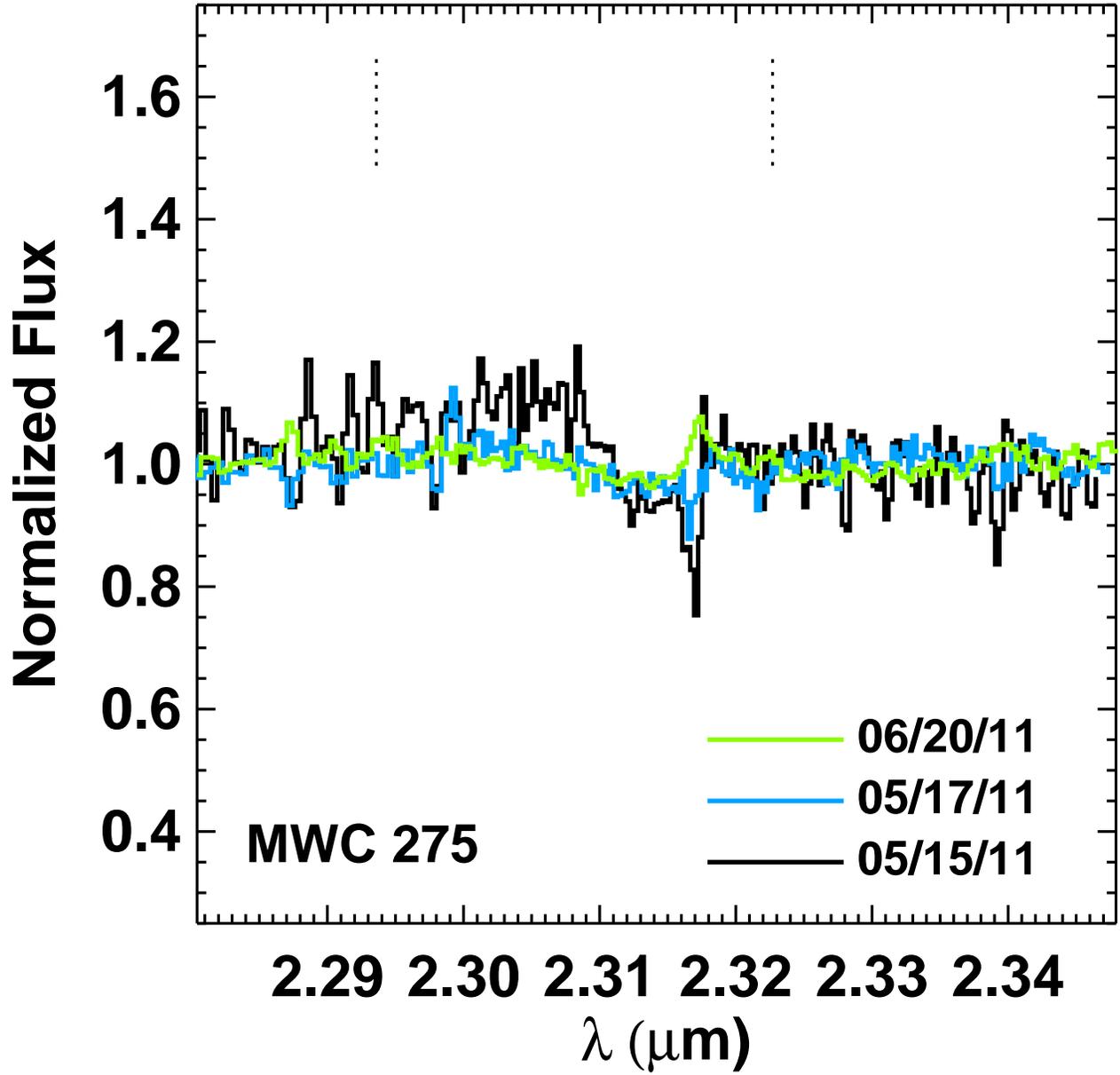}
\caption{Spectra of MWC 275.
\label{fig:mwc275}}
\end{figure}

\begin{figure}
\plotone{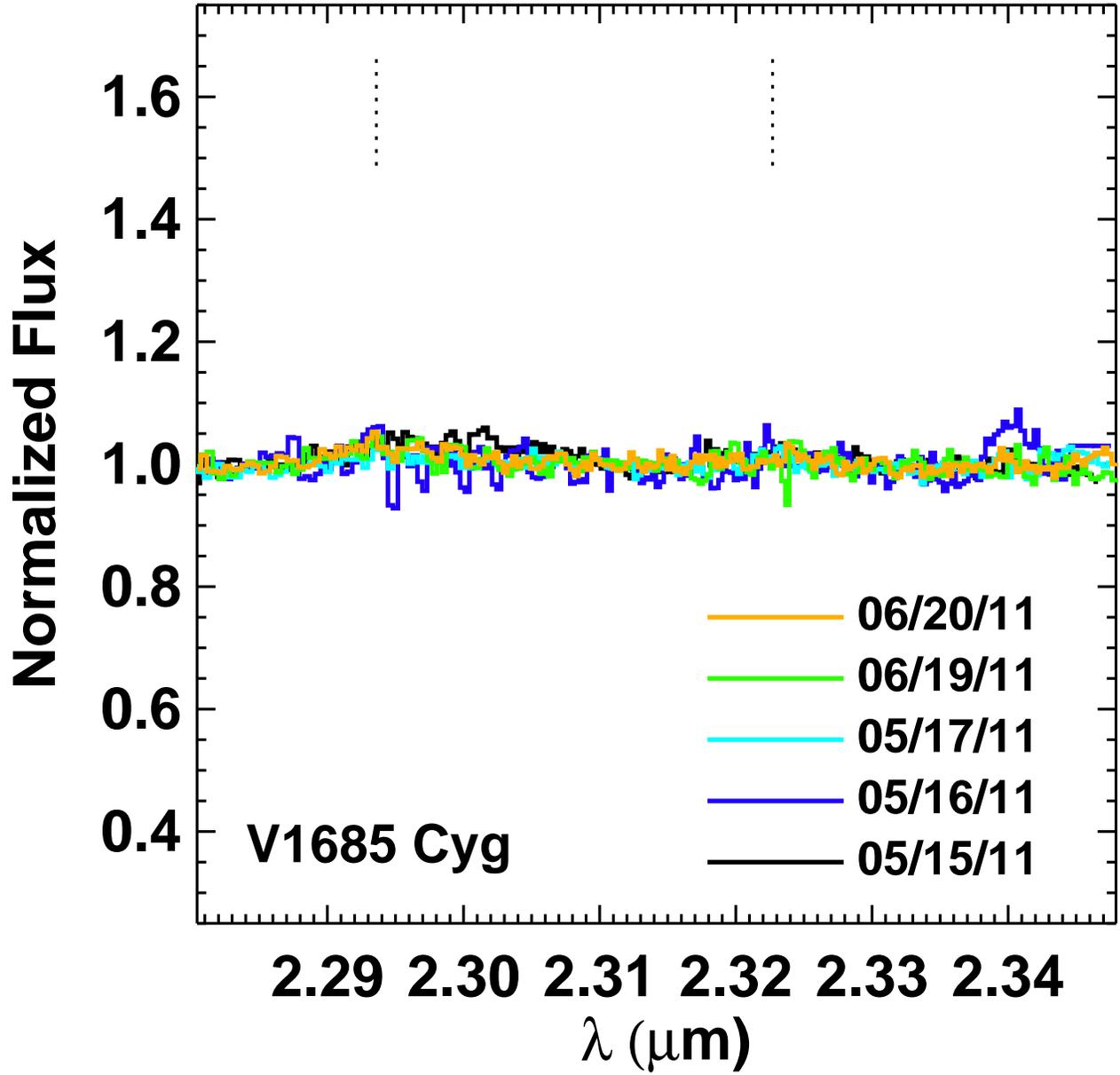}
\caption{Spectra of V1685 Cyg.
\label{fig:v1685}}
\end{figure}

\begin{figure}
\plotone{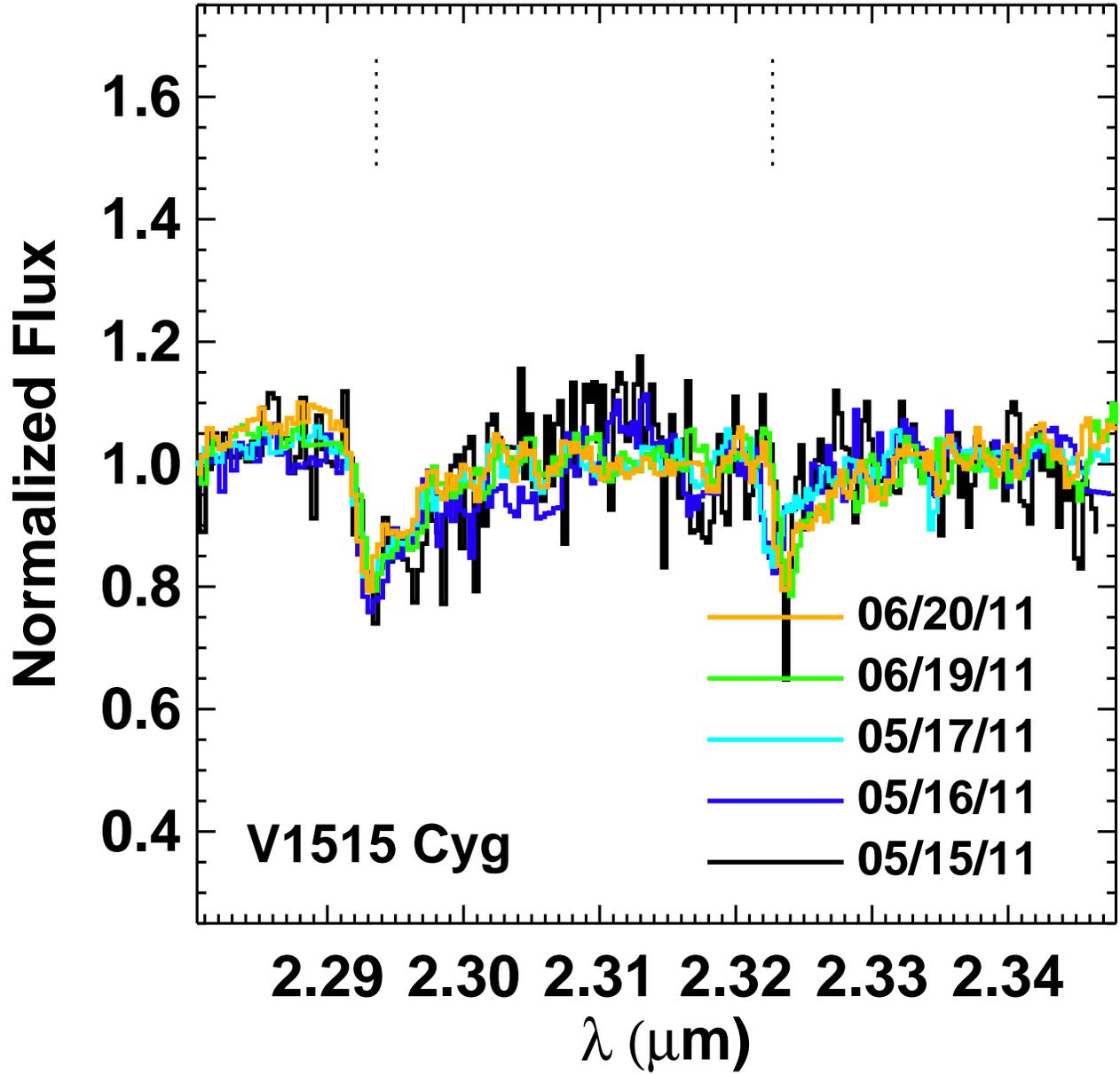}
\caption{Spectra of V1515 Cyg.
\label{fig:v1515}}
\end{figure}

\begin{figure}
\plotone{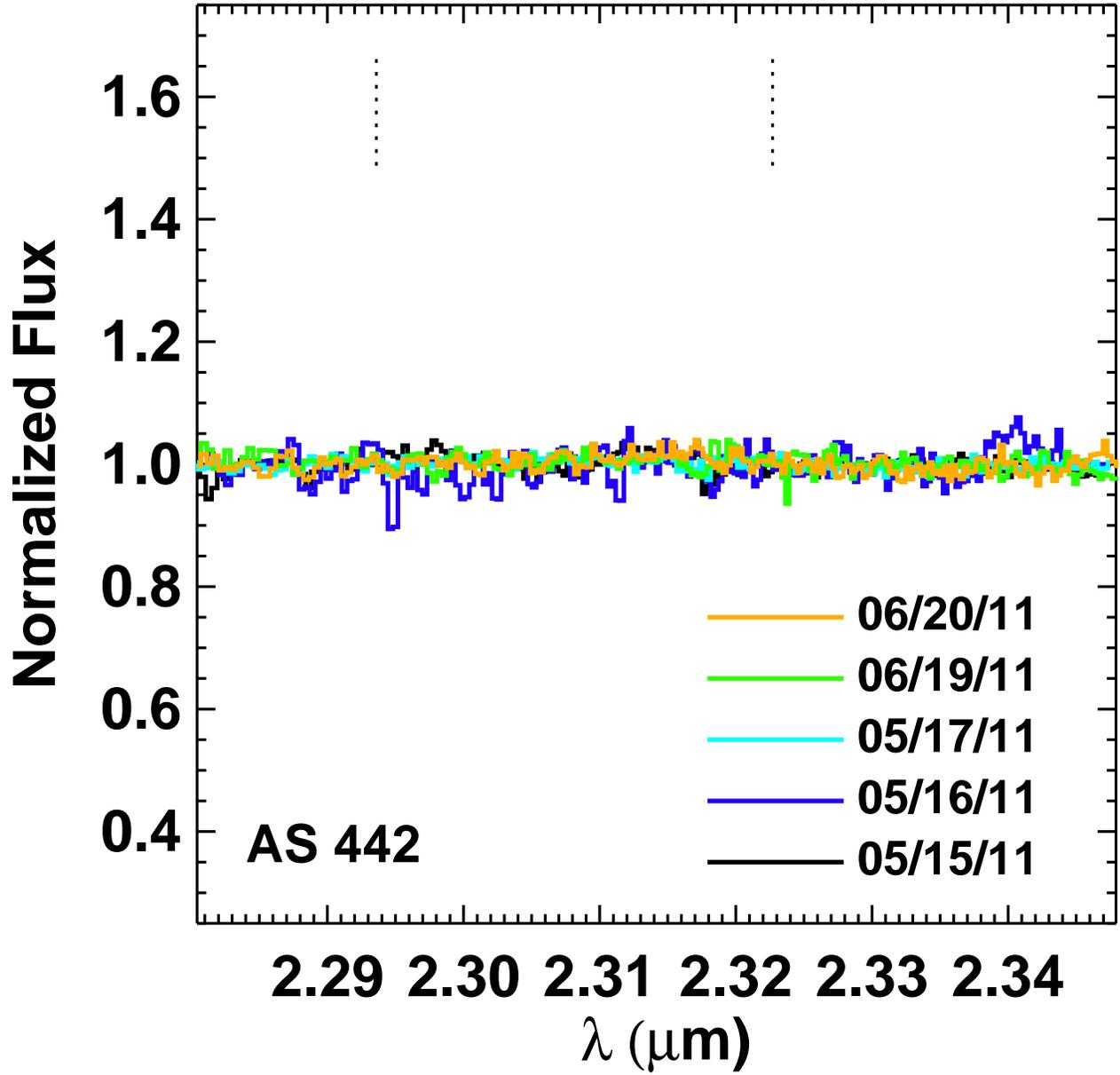}
\caption{Spectra of AS 442.
\label{fig:as442}}
\end{figure}

\begin{figure}
\plotone{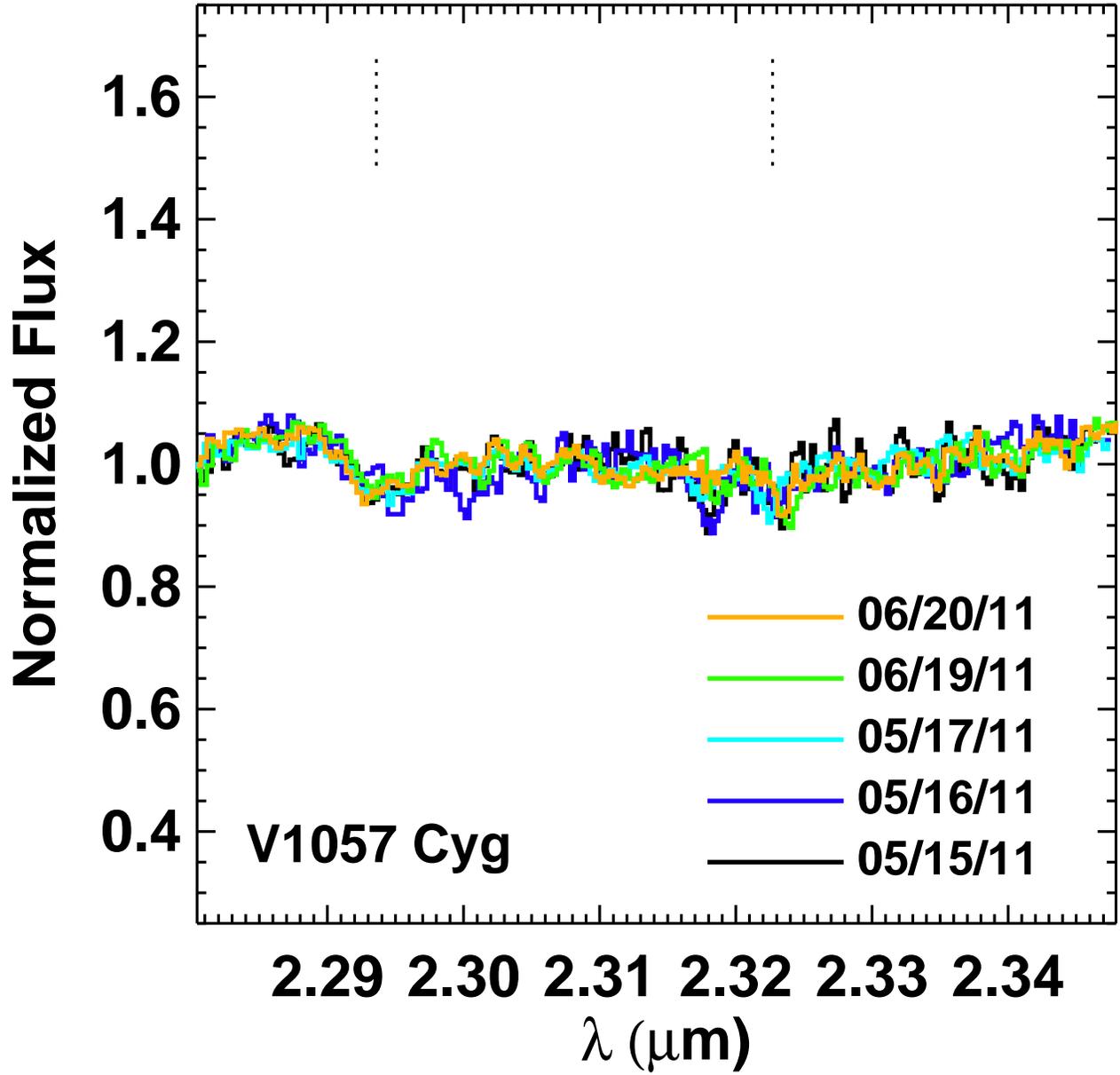}
\caption{Spectra of V1057 Cyg.
\label{fig:v1057}}
\end{figure}

\begin{figure}
\plotone{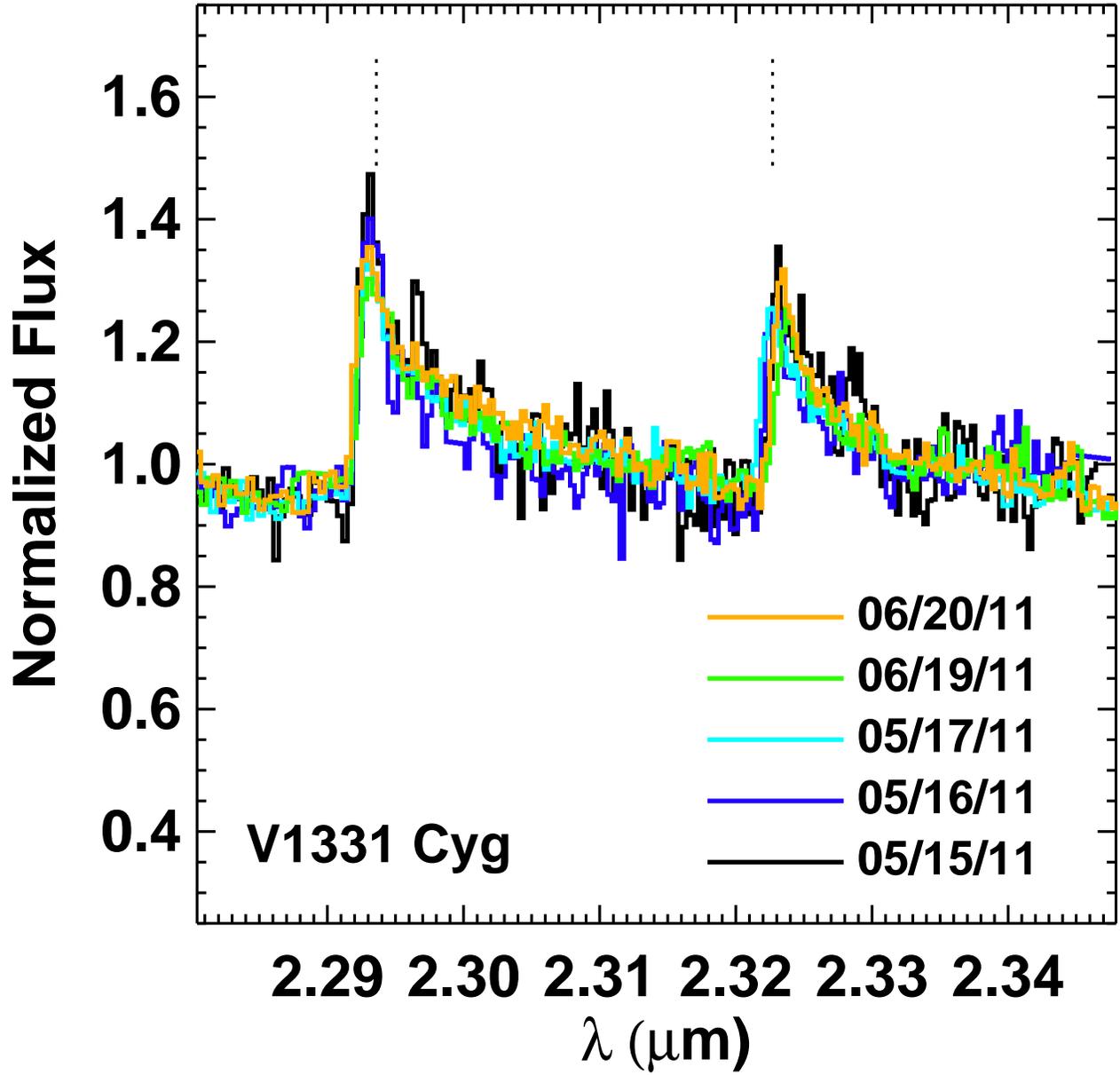}
\caption{Spectra of V1331 Cyg.
\label{fig:v1331}}
\end{figure}

Spectra of our targets are plotted in Figures
\ref{fig:as205}--\ref{fig:v1331}.  Many of our targets show no
evidence of CO emission or absorption in any observed epochs.  Some
targets show strong emission, and some show strong absorption
features.  However, as seen in Figures
\ref{fig:as205}--\ref{fig:v1331}, strong variability is generally
absent. While spectra do show differences from epoch-to-epoch, much or
all of this may be attributed to noise in the data.  

Beyond random noise, several objects (V2508 Oph, MWC 863, 51 Oph, MWC
275) appear to have variable features around 2.317 $\mu$m.  All of
these sources used a common telluric calibrator star.  The presence of
this feature in all of these objects suggests it to be an artifact of
calibration. There is a telluric absorption feature at approximately
this wavelength, which may also lead to imperfect calibration.


Because we do not calibrate the continuum flux level of our
observations, we can not calculate line fluxes in the  CO bandheads.
Rather we determine equivalent widths (EWs), which provide a
measurement of line flux relative to the continuum level.  To first
approximation, the continuum emission in the near-IR traces dust, and
so EWs provide a measure of CO gas to dust emission.  However,
some fraction of inner disk continuum emission may trace gas, perhaps
free-free emission from H or H$^{-}$
\citep[e.g.,][]{TANNIRKULAM+08,EISNER+09}.

We measure equivalent widths (EWs) of the
$v=2\rightarrow 0$ and $v=3\rightarrow 1$ CO bandhead features.  After
subtracting the continuum level,
EWs are measured by
simply integrating the flux in spectral regions appropriate to each
bandhead.  For the $v=2\rightarrow 0$ bandhead we integrate from 2.290
to 2.317 $\mu$m.  For the $v=3\rightarrow 1$ bandhead we integrate
from 2.320 to 2.346 $\mu$m.  

To estimate errors in measured EWs, we simulate 1000 noise
realizations for each spectrum.  The noise is Gaussian, with $\sigma$
determined from line-free regions of the observed spectrum.  In
addition, we include the effects of bad pixels at random locations,
multiplying two pixel values in each synthetic spectrum by 5.  
The uncertainty in the data is estimated as the 3$\sigma$ confidence
level of the 1000 resulting synthetic spectra.  We choose
3$\sigma$ to facilitate inclusion of bad pixel events in our
estimated errors. EWs and uncertainties for each target and observed epoch
are listed in Table \ref{tab:ews}.

For essentially all sources in our sample, the EWs do not indicate
statistically significant variations across epochs.  As listed in
Table \ref{tab:ews}, 51 Oph appears as if it may be an exception, with
variations in the $v = 2 \rightarrow 0$ bandhead emission.  However
inspection of Figure \ref{fig:51oph} suggests that the epoch with a
larger inferred EW may have a hot pixel coincident with the peak of
the bandhead, and that the EW may thus be contaminated.  The fact that
the $v = 3 \rightarrow 1$ bandhead does not display variability
supports this hypothesis.

\section{DISCUSSION \label{sec:disc}}
Previous studies have shown the CO overtone bandhead emission or
absorption to be variable in a number of objects
\citep[e.g.,][]{BISCAYA+97,NAJITA+00,NCM03}.  DG Tau, in particular,
is seen to vary significantly, with CO emission strong in some epochs
\citep[e.g.,][]{CARR89, BISCAYA+97} and undetected in others
\citep[e.g.,][]{GL96,BISCAYA+97,EISNER+09}.  While our sample does not
include DG Tau, several of our sample
objects have been observed previously, allowing comparison between the
results presented above and conclusions from the literature.

For example, in observations of V1331 Cyg across several nights and
months, the EW of the $v=2\rightarrow 0$ CO bandhead emission varied
by nearly a factor of two \citep{BISCAYA+97}.
The largest variability was observed over a one
week interval, while smaller changes were observed over month-to-month
timescales.  In contrast our spectra of V1331 Cyg, which also sample
day-to-day and month-to-month timescales, show no significant
variability in either the  $v=2\rightarrow 0$  or  $v=3\rightarrow 1$
bandheads (Table \ref{tab:ews}).  

Our spectra, while approximately
constant across the observed epochs, do differ from previous
observations.  The EWs we measure for the $v = 2 \rightarrow 0$
bandhead appear to be $\la 50\%$ those measured by \citet{BISCAYA+97}.
Other previous observations of V1331 Cyg, generally with different
spectral resolution than those presented here, also suggest higher EWs
than seen in our data \citep[e.g.,][]{NAJITA+09}.  Evidently the CO
bandheads may vary sometimes, even though we do not see variations
during the time-baseline of our observations.

Photometric observations of V1331 Cyg  in the near-IR ($JHK$) suggest a
similar trend. No significant variability in the flux of  V1331 Cyg
was observed on timescales of days, months,
or years \citep{EISNER+07c}.  However differences between
measurements in the literature, spanning decades, do suggest variations by up to
$\sim 0.5$ magnitudes
\citep{MENDOZA68,GP74,CK79,SUN+91,CUTRI+03,ABRAHAM+04}.  
These observations imply that V1331 Cyg has spent significant periods
of time in at least two distinct photometric states.  Thus, this object may have
episodic variations on timescales not sampled in monitoring
observations \citep[e.g.,][]{EISNER+07c}.

Multi-epoch observations of V1057 Cyg also suggested variability, in
this case in CO overtone absorption spectra \citep{BISCAYA+97, HK87}.
However this variability was inferred between spectra taken with
different telescopes, and indeed very different spectrographs (one was
an FTS spectrum and one a grating spectrum).  In our observations of
this object no significant variability is seen across days or months.
While the EW of the $v=2 \rightarrow 0$ bandhead measured here is
lower than that measured by \citet{BISCAYA+97}, it is in fact
consistent with the older measurement of \citet{HK87}.

\citet{BISCAYA+97} suggested that they may have seen V1057 Cyg in
outburst during their observations; this suggestion seems consistent
with our observations and the older data from \citet{HK87} tracing the
quiescent state.  As noted above, CO absorption in V1057 Cyg (and other
FU Ori stars) likely traces circumstellar disk material rather than a stellar
photosphere.  The CO bandhead shapes for V1057 Cyg \citep[Figure
\ref{fig:v1057}; see also][]{BISCAYA+97} differ from stellar
photospheric absorption features, supporting this conclusion.  The
variability inferred between different observations in the literature
likely arises in a circumstellar disk.

While our observations of 51 Oph suggested a higher EW in one
of the CO bandheads in one epoch (Table \ref{tab:ews}), we argued
above that this is probably due to an uncorrected bad pixel rather
than true variability.  Moreover, previous observations \citep[at higher
spectral resolution;][]{THI+05,BERTHOUD+07} show EWs for the $v = 2
\rightarrow 0$ bandhead that are consistent with those measured here.
Both \citet{THI+05} and \citet{BERTHOUD+07} find that the EWs
of the   $v = 3 \rightarrow 1$ and   $v = 2 \rightarrow 0$ bandheads
are comparable, while our results suggest the EWs of the $v = 3 \rightarrow 1$
bandhead are smaller.  This may indicate long-term variation in the
temperature of the CO emission, which would affect the ratio of the
two bandheads (see Figure \ref{fig:models} below).  However the
difference between the EWs of the two bandheads in our data is only
marginally significant (Table \ref{tab:ews}). 

None of our sample---including 12 young star+disk
systems---exhibit significant variability in their CO overtone
emission/absorption during the observed epochs (Table \ref{tab:ews}).
The fact that other objects, including some in our sample, have been
seen to vary in previous observations begs the question of why we do
not see such variability.

One possibility is that the duty cycle of variability in CO overtone
emission or absorption features is low.  For this scenario, we may
simply have missed variability during the $\sim 1$ month of
observations presented here.  However under this scenario, multiple
previous observations (some of similar duration to the observations
presented here) would have needed to be
lucky to have seen variations.

Another potential explanation is that some of the previous inferred
variability was actually due to variable calibration, rather than
intrinsic variability.  Given that some of the previous observations also
used the FSPEC instrument \citep{BISCAYA+97}, this seems
unlikely. Moreover, previous large-amplitude variability observed at
the MMT \citep[also with FSPEC;][]{BISCAYA+97} is difficult to explain through
instrumental effects.  However comparison of different datasets,
taken with different instruments, may suffer to some extent from a
lack of consistent calibration.

In terms of physical implications, the lack of significant variability
in our data indicates a relatively quiescent inner disk environment
around our sample objects.  
The magnitude of variability that is ruled out corresponds to ranges
of gas temperatures or surface densities.  For optically thin gas, the
emission intensity scales directly with density, and thus the
uncertainties in our data correspond directly to the amount of allowed
variations in gas density.  Including some uncertainty on estimation
of continuum levels (which may influence the total measured EW), our
data suggest that gas densities do not vary by more than $\sim 20\%$
during the observed epochs.  Temperature variations may affect EWs,
and will also change the relative strengths of the two CO overtone
bandheads observed here.  Lack of significant variations in either EWs
of individual features or the ratios of EWs for the two bandheads
indicates that gas temperature does not vary by more than $\sim
20\%$.  

To illustrate how variable densities and temperatures affect the CO
overtone bandheads, we generate synthetic spectra for different
parameter values.  Using CO opacities from the HITRAN/HITEMP database
\citep{ROTHMAN+05}, we generate synthetic spectra, then smooth them to
the resolution of our data.  (We add a continuum level of unity to
these spectra, to match the normalized spectra presented above.)
Example spectra where density or
temperature has been varied by $\sim 20\%$ are shown in Figure
\ref{fig:models}. Comparison of this Figure with the spectra of our
observed sources shows that such variations--or anything larger--would
produce observable effects.  Indeed, the EWs and ratios of EWs for the
two overtone bandheads in these examples vary more than seen in any of
our sample objects.

If inner disk regions varied on dynamical
timescales, we would expect to see variations in the CO
emission/absorption on timescales of $\tau_{\rm dyn} =
\sqrt{R^3/GM_{\ast}}$.   Assuming that CO gas is found between
magnetospheric and dust sublimation radii, measured radii for
protoplanetary disks in our sample
\citep{EISNER+04,EISNER+05,EISNER+07c,TATULLI+08,EH11} imply dynamical
times $\la 10$ days.  
Similar variability timescales are expected for processes related to
the interaction between stellar magnetospheres and inner accretion
disks.  Changes in the magnetospheric radius with time
\citep[e.g.,][]{GBW99,GW99}, or the intrusion of
time-variable accretion streams through the magnetosphere
\citep[e.g.,][]{KR09}, can alter the inner gaseous disk structure
on timescales similar to $\tau_{\rm dyn}$.  

The lack of observed variations on
these timescales, which are well-sampled with our observing cadence,
suggests a lack of inner disk changes on dynamical timescales.  We
note, though, that one of our targets did show such variability in a
previous study \citep{BISCAYA+97}.

Variation in the CO overtone emission may also occur because of
density  or temperature changes on timescales longer than those
sampled here.  The density of gas may vary if accretion through the
disk is not steady; such variations would occur on viscous
timescales.  For an $\alpha$ disk \citep[e.g.,][]{SS73}, 
viscous timescales are 
\begin{displaymath}\
\tau_{\rm visc} \sim \frac{\tau_{\rm dyn}}{\alpha} \left(\frac{R}{H}\right)^2.
\end{displaymath}
Typical values for the ratio of disk height to radius ($H/R \sim 0.1$) and
$\alpha \sim 0.01$ yield inner disk viscous timescales of $\ga 100$ years for
our sample.  Variable gas temperature, for example due to changing
heating from a variable central star, can also lead to CO
variability.  Such variability would occur on thermal timescales,
$\tau_{\rm therm} \sim \tau_{\rm dyn} / \alpha \ga 1$ year for our sample.
While our data do not sample such timescales, long-term variations
between different observations in the literature may be due, in part,
to such effects.

\section{Conclusions}
We presented multi-epoch FSPEC observations of the CO overtone bandheads
in a sample of young stars.  CO emission or absorption was observed
in a number of targets, although no significant variability of the
emission/absorption features was seen across timescales of days or
months.  In contrast, previous monitoring observations have found
variability on such timescales.  Moreover, for sources with previous
CO-band spectra in the literature, the features seen in our
data do appear different than those seen in previous observations.

These results indicate that variability in the CO overtone bandheads
may have a low duty cycle, or that the sources in our sample undergo
significant quiescent periods when the variability is small.  
Our observations sampled well the inner disk dynamical timescales of
our targets, indicating that the inner disk density and/or temperature
did not vary at a level above $\sim 20 \%$. 
The origin of long-term variations--i.e.,
those between our observations and previous data in the literature--is
less well-constrained.

\noindent{\bf Acknowledgments.}
JAE gratefully acknowledges support from an Alfred P. Sloan Research
Fellowship.  Support for this work, largely for students affiliated
with the project, comes from the National Science
Foundation under Award No. AST-0847170, a PAARE Grant for the
California-Arizona Minority Partnership for Astronomy Research and
Education (CAMPARE).   Any opinions, findings, and conclusions or
recommendations expressed in this material are those of the author(s)
and do not necessarily reflect the views of the National Science
Foundation.

\clearpage

\begin{deluxetable}{lccccc|ccc}
\tabletypesize{\scriptsize}
\tablewidth{0pt}
\tablecaption{Observed Targets and Epochs
\label{tab:sample}}
\tablehead{\colhead{Source} & \colhead{$\alpha$} & \colhead{$\delta$}
  & \colhead{Night} & \colhead{$T_{\rm
      int}$} & \colhead{Airmass} & \colhead{Calibrators}
  &\colhead{$T_{\rm int}$} & \colhead{Airmass}  \\
 & (J2000) & (J2000) & & (s) & &  & (s) & }
\startdata
AS 205 N & 16 11 31.40 & -18 38 24.5 & 05/15/11 & 300 & 2.1 & HD 145127 & 150, 300 &
2.4, 2.0 \\
 & & & 05/17/11 & 300 & 1.7 & HD 145188 & 300 & 1.9 \\
 & & & 06/20/11 & 125 & 1.7 & HD 145188 & 250 & 1.8 \\
V1002 Sco & 16 12 40.51 & -18 59 28.1 & 05/15/11 & 150 & 2.0 & HD 145127 & 150,300 &
2.4, 2.0  \\
 & & & 05/17/11 & 300 & 1.7 & HD 145188 & 300 & 1.9 \\
 & & & 06/20/11 & 250 & 1.7 & HD 145188 & 250 & 1.8 \\
V2508 Oph & 16 48 45.63 & -14 16 40.0 & 05/15/11 & 300 & 2.0 & HD 145127 & 150, 300 &
2.4, 2.0 \\
 & & & 05/17/11 & 300 & 1.6 & HD 145188, HD 145127 & 300, 150 & 1.9, 1.8 \\
 & & & 06/20/11 & 250 & 1.5 & HD 145188 & 250 & 1.8 \\
AS 209 & 16 49 15.30 & -14 22 08.6 & 05/15/11 & 300 & 1.9 & HD 145127 & 150, 300 &
2.4, 2.0 \\
& & & 05/17/11 & 300 & 1.5 & HD 145127 & 150 & 1.8 \\
 & & & 06/20/11 & 250 & 1.5 & HD 145188 & 250 & 1.8 \\
MWC 863 & 16 40 17.92 & -23 53 45.2 &  05/15/11 & 150 & 2.1 & HD 145127 & 300 & 2.0 \\
& & & 05/17/11 & 150 & 1.8 & HD 145127 & 150 & 1.8 \\
& & & 06/20/11 & 125 & 2.0 & HD 145127 & 250 & 2.2 \\
51 Oph & 17 31 24.95 & -23 57 45.5 & 05/15/11 & 150 & 2.3 & HD 145127 & 300 & 2.0  \\
& & & 05/17/11 & 150 & 1.8 & HD 145127 & 150 & 1.8 \\
& & & 06/20/11 & 125 & 1.9 & HD 145127 & 250 & 2.2 \\
MWC 275 & 17 56 21.29 & -21 57 21.9 & 05/15/11 & 150 & 2.4 & HD 145127 & 300 & 2.0 \\
& & & 05/17/11 & 150 & 1.8 & HD 145127 & 150 & 1.8 \\
& & & 06/20/11 & 125 & 1.7 & HD 145127 & 250 & 2.2 \\
V1685 Cyg & 20 20 28.24 & 41 21 51.6 & 05/15/11 & 300 & 1.4 & HD 192538 & 300 & 1.5 \\
 & & & 05/16/11 & 150 & 1.3 & HD 192538, HD 199312 & 300, 450 & 1.3, 1.3 \\
 & & & 05/17/11 & 300 & 1.1 & HD 192538, HD 199312 & 300, 300 & 1.1, 1.1 \\
 & & & 06/19/11 & 250 & 1.0 & HD 192538, HD 199312 & 250, 375 & 1.0, 1.0 \\
 & & & 06/20/11 & 125 & 1.0 & HD 192538, HD 199312 & 125, 250 & 1.0, 1.1 \\
V1515 Cyg & 20 23 48.02 & 42 12 25.8 &  05/15/11 & 300 & 1.6 & HD 192538 & 300 & 1.5 \\ 
 & & & 05/16/11 & 570 & 1.5 & HD 192538 & 300 & 1.3 \\
 & & & 05/17/11 & 300 & 1.2 & HD 192538 & 300 & 1.1 \\
 & & & 06/19/11 & 375 & 1.0 & HD 192538 & 250 & 1.0 \\
 & & & 06/20/11 & 250 & 1.1 & HD 192538 & 125 & 1.0 \\
AS 442 & 20 47  37.47 & 43 47 24.9 & 05/15/11 &  300 & 1.5 & HD 192538 & 300 & 1.5\\
 & & & 05/16/11 & 300 & 1.3 & HD 192538, HD 199312 & 300, 450 & 1.3, 1.3 \\
 & & & 05/17/11 & 300 & 1.1 & HD 192538, HD 199312 & 300, 300 & 1.1, 1.1 \\
 & & & 06/19/11 & 125 & 1.0 & HD 192538, HD 199312 & 250, 375 & 1.0, 1.0 \\
 & & & 06/20/11 & 125 & 1.1 & HD 192538, HD 199312 & 125, 250 & 1.0, 1.1 \\
V1057 Cyg & 20 58 53.73 & 44 15 28.5 & 05/15/11 & 300 & 1.8 & HD 192538 & 300 & 1.5 \\
 & & & 05/16/11 & 300 & 1.6 & HD 192538, HD 199312 & 300, 450 & 1.3, 1.3 \\
 & & & 05/17/11 & 300 & 1.2 & HD 192538 & 300 & 1.1 \\
 & & & 06/19/11 & 200 & 1.0 & HD 192538 & 250 & 1.0 \\
 & & & 06/20/11 & 125 & 1.1 & HD 192538 & 125 & 1.0 \\
V1331 Cyg & 21 01 09.21 & 50 21 44.8 &  05/15/11 & 150 & 1.7 &  HD 192538 & 300 & 1.5 \\
&  & & 05/16/11 & 150 & 1.3 & HD 193369, HD 199312 & 150, 450 & 1.1,
1.3 \\
 & & & 05/17/11 & 150 & 1.1 & HD 192538 & 300 & 1.1 \\
& & & 06/19/11 & 500 & 1.1 & HD 192538 & 250 & 1.0  \\
& & & 06/20/11 &  250 & 1.1 & HD 192538 & 125 & 1.0 \\
\enddata
\tablecomments{Sources are listed roughly in order or right ascension,
  keeping targets that share common calibration in consecutive order.}
\end{deluxetable}

\begin{deluxetable}{lccc}
\tabletypesize{\scriptsize}
\tablewidth{0pt}
\tablecaption{CO Bandhead Equivalent Widths
\label{tab:ews}}
\tablehead{\colhead{Source} &\colhead{Night}  & \colhead{EW} &
  \colhead{EW} \\
 & &  \colhead{$v=2 \rightarrow 0$} & \colhead{$v=3 \rightarrow 1$} \\
 & & \colhead{(\AA)} & \colhead{(\AA)}}
\startdata
AS 205N & 05/15/11 &  $-1 \pm  1$ & $  2 \pm  1$ \\
 & 05/17/11 & $  0 \pm  1$ & $  1 \pm  1$ \\
 & 06/20/11 & $ -3 \pm  2$ & $  2 \pm  2$ \\
V1002 Sco & 05/15/11 & $ 12 \pm  3$ & $ 18 \pm  3$ \\
 & 05/17/11 & $ 13 \pm  3$ & $ 16 \pm  3$ \\
 & 06/20/11 & $ 10 \pm  4$ & $ 18 \pm  4$ \\
V2508 Oph & 05/15/11 & $  9 \pm  4$ & $ 19 \pm  4$ \\
 & 05/17/11 & $ 17 \pm  2$ & $ 16 \pm  2$ \\
 & 06/20/11 & $ 14 \pm  3$ & $ 20 \pm  3$ \\
AS 209 & 05/15/11 & $  9 \pm  2$ & $ 14 \pm  2$ \\
 & 05/17/11 & $ 12 \pm  2$ & $ 12 \pm  2$ \\
 & 06/20/11 & $  8 \pm  3$ & $ 15 \pm  3$ \\
MWC 863 & 05/15/11 & $  1 \pm  3$ & $  9 \pm  3$ \\
 & 05/17/11 & $  5 \pm  3$ & $  7 \pm  2$ \\
 & 06/20/11 & $  6 \pm  1$ & $  9 \pm  1$ \\
51 Oph & 05/15/11 & $-32 \pm  2$ & $-18 \pm  2$ \\
 & 05/17/11 & $-23 \pm  3$ & $-19 \pm  2$ \\
 & 06/20/11 & $-23 \pm  3$ & $-16 \pm  3$ \\
MWC 275 & 05/15/11 & $  2 \pm  6$ & $ 14 \pm  6$ \\
 & 05/17/11 & $ 13 \pm  2$ & $ 11 \pm  2$ \\
 & 06/20/11 & $ 11 \pm  2$ & $ 12 \pm  2$ \\
V1685 Cyg & 05/15/11 & $ -2 \pm  1$ & $  1 \pm  1$ \\
 & 05/16/11 & $  2 \pm  2$ & $  0 \pm  2$ \\
 & 05/17/11 & $  1 \pm  1$ & $  1 \pm  1$ \\
 & 06/19/11 & $  0 \pm  2$ & $  2 \pm  1$ \\
 & 06/20/11 & $  0 \pm  1$ & $  2 \pm  1$ \\
V1515 Cyg & 05/15/11 & $ 18 \pm  7$ & $ 18 \pm  6$ \\
 & 05/16/11 & $ 25 \pm  3$ & $ 17 \pm  3$ \\
 & 05/17/11 & $ 23 \pm  3$ & $ 18 \pm  2$ \\
 & 06/19/11 & $ 21 \pm  3$ & $ 21 \pm  3$ \\
 & 06/20/11 & $ 21 \pm  4$ & $ 19 \pm  4$ \\
AS 442 & 05/15/11 & $  0 \pm  2$ & $  1 \pm  1$ \\
 & 05/16/11 & $  3 \pm  3$ & $  0 \pm  2$ \\
 & 05/17/11 & $  0 \pm  1$ & $  2 \pm  1$ \\
 & 06/19/11 & $  0 \pm  2$ & $  1 \pm  1$ \\
 & 06/20/11 & $  0 \pm  1$ & $  2 \pm  1$ \\
V1057 Cyg & 05/15/11 & $ 11 \pm  2$ & $ 10 \pm  2$ \\
 & 05/16/11 & $ 14 \pm  3$ & $ 10 \pm  3$ \\
 & 05/17/11 & $ 13 \pm  2$ & $ 10 \pm  2$ \\
 & 06/19/11 & $ 11 \pm  3$ & $ 12 \pm  3$ \\
 & 06/20/11 & $ 13 \pm  2$ & $ 11 \pm  2$ \\
V1331 Cyg & 05/15/11 & $-35 \pm  6$ & $-26 \pm  6$ \\
 & 05/16/11 & $-27 \pm  6$ & $-25 \pm  5$ \\
 & 05/17/11 & $-33 \pm  4$ & $-24 \pm  4$ \\
 & 06/19/11 & $-33 \pm  3$ & $-24 \pm  2$ \\
 & 06/20/11 & $-39 \pm  3$ & $-26 \pm  2$ \\
\enddata
\tablecomments{Uncertainties include statistical uncertainties and
  uncertainties related to uncorrected bad pixels, but do
  not include errors in estimation of continuum level.}
\end{deluxetable}

\clearpage

\begin{figure}
\plotone{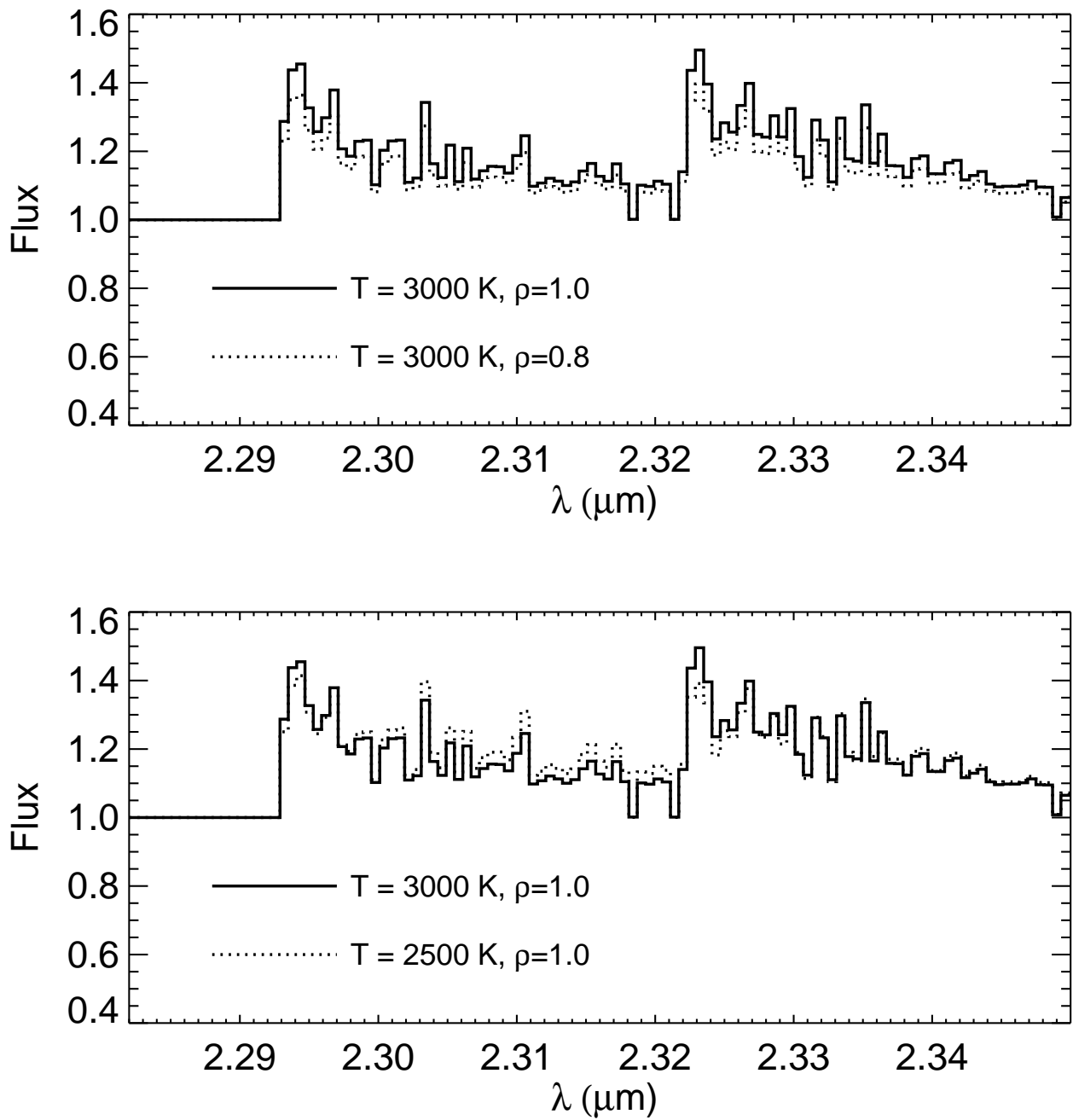}
\caption{Synthetic CO spectra for different values of gas density or
  temperature.  We generated these spectra using HITRAN/HITEMP opacities,
  sampling at the resolution of the FSPEC data presented above.  CO
  spectra were then added to a continuum level of unity. The
  densities are chosen to produce line-to-continuum ratios consistent
  with those of observed targets, and $\rho = 1$ represents a
  baseline model with CO column density $N_{\rm CO} = 5 \times
  10^{21}$ cm$^{-2}$. 
\label{fig:models}}
\end{figure}


\begin{thebibliography}{45}
\expandafter\ifx\csname natexlab\endcsname\relax\def\natexlab#1{#1}\fi

\bibitem[{{{\'A}brah{\'a}m} {et~al.}(2004){{\'A}brah{\'a}m}, {K{\'o}sp{\'a}l},
  {Csizmadia}, {Kun}, {Mo{\'o}r}, \& {Prusti}}]{ABRAHAM+04}
{{\'A}brah{\'a}m}, P., {K{\'o}sp{\'a}l}, {\'A}., {Csizmadia}, S., {Kun}, M.,
  {Mo{\'o}r}, A., \& {Prusti}, T. 2004, \aap, 428, 89

\bibitem[{{Berthoud}(2008)}]{BERTHOUD08}
{Berthoud}, M.~G. 2008, PhD thesis, Cornell University

\bibitem[{{Berthoud} {et~al.}(2007){Berthoud}, {Keller}, {Herter}, {Richter},
  \& {Whelan}}]{BERTHOUD+07}
{Berthoud}, M.~G., {Keller}, L.~D., {Herter}, T.~L., {Richter}, M.~J., \&
  {Whelan}, D.~G. 2007, \apj, 660, 461

\bibitem[{{Biscaya} {et~al.}(1997){Biscaya}, {Rieke}, {Narayanan}, {Luhman}, \&
  {Young}}]{BISCAYA+97}
{Biscaya}, A.~M., {Rieke}, G.~H., {Narayanan}, G., {Luhman}, K.~L., \& {Young},
  E.~T. 1997, \apj, 491, 359

\bibitem[{{Brittain} {et~al.}(2007){Brittain}, {Simon}, {Najita}, \&
  {Rettig}}]{BRITTAIN+07}
{Brittain}, S.~D., {Simon}, T., {Najita}, J.~R., \& {Rettig}, T.~W. 2007, \apj,
  659, 685

\bibitem[{{Calvet} {et~al.}(1991){Calvet}, {Patino}, {Magris}, \&
  {D'Alessio}}]{CALVET+91}
{Calvet}, N., {Patino}, A., {Magris}, G.~C., \& {D'Alessio}, P. 1991, \apj,
  380, 617

\bibitem[{{Carpenter} {et~al.}(2001){Carpenter}, {Hillenbrand}, \&
  {Skrutskie}}]{CHS01}
{Carpenter}, J.~M., {Hillenbrand}, L.~A., \& {Skrutskie}, M.~F. 2001, \aj, 121,
  3160

\bibitem[{{Carr}(1989)}]{CARR89}
{Carr}, J.~S. 1989, \apj, 345, 522

\bibitem[{{Cohen} \& {Kuhi}(1979)}]{CK79}
{Cohen}, M. \& {Kuhi}, L.~V. 1979, \apjs, 41, 743

\bibitem[{{Cutri} {et~al.}(2003){Cutri}, {Skrutskie}, {van Dyk}, {Beichman},
  {Carpenter}, {Chester}, {Cambresy}, {Evans}, {Fowler}, {Gizis}, {Howard},
  {Huchra}, {Jarrett}, {Kopan}, {Kirkpatrick}, {Light}, {Marsh}, {McCallon},
  {Schneider}, {Stiening}, {Sykes}, {Weinberg}, {Wheaton}, {Wheelock}, \&
  {Zacarias}}]{CUTRI+03}
{Cutri}, R.~M., {Skrutskie}, M.~F., {van Dyk}, S., {Beichman}, C.~A.,
  {Carpenter}, J.~M., {Chester}, T., {Cambresy}, L., {Evans}, T., {Fowler}, J.,
  {Gizis}, J., {Howard}, E., {Huchra}, J., {Jarrett}, T., {Kopan}, E.~L.,
  {Kirkpatrick}, J.~D., {Light}, R.~M., {Marsh}, K.~A., {McCallon}, H.,
  {Schneider}, S., {Stiening}, R., {Sykes}, M., {Weinberg}, M., {Wheaton},
  W.~A., {Wheelock}, S., \& {Zacarias}, N. 2003, {2MASS All Sky Catalog of
  point sources.} (The IRSA 2MASS All-Sky Point Source Catalog, NASA/IPAC
  Infrared Science Archive.~http://irsa.ipac.caltech.edu/applications/Gator/)

\bibitem[{{Eiroa} {et~al.}(2002){Eiroa}, {Oudmaijer}, {Davies}, {de Winter},
  {Garz{\' o}n}, {Palacios}, {Alberdi}, {Ferlet}, {Grady}, {Cameron}, {Deeg},
  {Harris}, {Horne}, {Mer{\'{\i}}n}, {Miranda}, {Montesinos}, {Mora}, {Penny},
  {Quirrenbach}, {Rauer}, {Schneider}, {Solano}, {Tsapras}, \&
  {Wesselius}}]{EIROA+02}
{Eiroa}, C., {Oudmaijer}, R.~D., {Davies}, J.~K., {de Winter}, D., {Garz{\'
  o}n}, F., {Palacios}, J., {Alberdi}, A., {Ferlet}, R., {Grady}, C.~A.,
  {Cameron}, A., {Deeg}, H.~J., {Harris}, A.~W., {Horne}, K., {Mer{\'{\i}}n},
  B., {Miranda}, L.~F., {Montesinos}, B., {Mora}, A., {Penny}, A.,
  {Quirrenbach}, A., {Rauer}, H., {Schneider}, J., {Solano}, E., {Tsapras}, Y.,
  \& {Wesselius}, P.~R. 2002, \aap, 384, 1038

\bibitem[{{Eisner} {et~al.}(2010){Eisner}, {Doppmann}, {Najita}, {McCarthy},
  {Kulesa}, {Swift}, \& {Teske}}]{EISNER+10b}
{Eisner}, J.~A., {Doppmann}, G.~W., {Najita}, J.~R., {McCarthy}, D., {Kulesa},
  C., {Swift}, B.~J., \& {Teske}, J. 2010, \apjl, 722, L28

\bibitem[{{Eisner} {et~al.}(2009){Eisner}, {Graham}, {Akeson}, \&
  {Najita}}]{EISNER+09}
{Eisner}, J.~A., {Graham}, J.~R., {Akeson}, R.~L., \& {Najita}, J. 2009, \apj,
  692, 309

\bibitem[{{Eisner} \& {Hillenbrand}(2011)}]{EH11}
{Eisner}, J.~A. \& {Hillenbrand}, L.~A. 2011, \apj, 738, 9

\bibitem[{{Eisner} {et~al.}(2005){Eisner}, {Hillenbrand}, {White}, {Akeson}, \&
  {Sargent}}]{EISNER+05}
{Eisner}, J.~A., {Hillenbrand}, L.~A., {White}, R.~J., {Akeson}, R.~L., \&
  {Sargent}, A.~I. 2005, \apj, 623, 952

\bibitem[{{Eisner} {et~al.}(2007){Eisner}, {Hillenbrand}, {White}, {Bloom},
  {Akeson}, \& {Blake}}]{EISNER+07c}
{Eisner}, J.~A., {Hillenbrand}, L.~A., {White}, R.~J., {Bloom}, J.~S.,
  {Akeson}, R.~L., \& {Blake}, C.~H. 2007, \apj, 669, 1072

\bibitem[{{Eisner} {et~al.}(2004){Eisner}, {Lane}, {Hillenbrand}, {Akeson}, \&
  {Sargent}}]{EISNER+04}
{Eisner}, J.~A., {Lane}, B.~F., {Hillenbrand}, L., {Akeson}, R., \& {Sargent},
  A. 2004, \apj, 613, 1049

\bibitem[{{Flaherty} {et~al.}(2012){Flaherty}, {Muzerolle}, {Rieke},
  {Gutermuth}, {Balog}, {Herbst}, {Megeath}, \& {Kun}}]{FLAHERTY+12}
{Flaherty}, K.~M., {Muzerolle}, J., {Rieke}, G., {Gutermuth}, R., {Balog}, Z.,
  {Herbst}, W., {Megeath}, S.~T., \& {Kun}, M. 2012, \apj, 748, 71

\bibitem[{{Forbrich} {et~al.}(2007){Forbrich}, {Preibisch}, {Menten},
  {Neuh{\"a}user}, {Walter}, {Tamura}, {Matsunaga}, {Kusakabe}, {Nakajima},
  {Brandeker}, {Fornasier}, {Posselt}, {Tachihara}, \& {Broeg}}]{FORBICH+07}
{Forbrich}, J., {Preibisch}, T., {Menten}, K.~M., {Neuh{\"a}user}, R.,
  {Walter}, F.~M., {Tamura}, M., {Matsunaga}, N., {Kusakabe}, N., {Nakajima},
  Y., {Brandeker}, A., {Fornasier}, S., {Posselt}, B., {Tachihara}, K., \&
  {Broeg}, C. 2007, \aap, 464, 1003

\bibitem[{{Glass} \& {Penston}(1974)}]{GP74}
{Glass}, I.~S. \& {Penston}, M.~V. 1974, \mnras, 167, 237

\bibitem[{{Goodson} {et~al.}(1999){Goodson}, {B{\"o}hm}, \& {Winglee}}]{GBW99}
{Goodson}, A.~P., {B{\"o}hm}, K.-H., \& {Winglee}, R.~M. 1999, \apj, 524, 142

\bibitem[{{Goodson} \& {Winglee}(1999)}]{GW99}
{Goodson}, A.~P. \& {Winglee}, R.~M. 1999, \apj, 524, 159

\bibitem[{{Greene} {et~al.}(2008){Greene}, {Aspin}, \& {Reipurth}}]{GAR08}
{Greene}, T.~P., {Aspin}, C., \& {Reipurth}, B. 2008, \aj, 135, 1421

\bibitem[{{Greene} \& {Lada}(1996)}]{GL96}
{Greene}, T.~P. \& {Lada}, C.~J. 1996, \aj, 112, 2184

\bibitem[{{Hartmann} {et~al.}(2004){Hartmann}, {Hinkle}, \& {Calvet}}]{HHC04}
{Hartmann}, L., {Hinkle}, K., \& {Calvet}, N. 2004, \apj, 609, 906

\bibitem[{{Hartmann} \& {Kenyon}(1985)}]{HK85}
{Hartmann}, L. \& {Kenyon}, S.~J. 1985, \apj, 299, 462

\bibitem[{{Hartmann} \& {Kenyon}(1987)}]{HK87}
---. 1987, \apj, 312, 243

\bibitem[{{Herbst} {et~al.}(1994){Herbst}, {Herbst}, {Grossman}, \&
  {Weinstein}}]{HERBST+94}
{Herbst}, W., {Herbst}, D.~K., {Grossman}, E.~J., \& {Weinstein}, D. 1994, \aj,
  108, 1906

\bibitem[{{Joy}(1942)}]{JOY42}
{Joy}, A.~H. 1942, \pasp, 54, 15

\bibitem[{{Kulkarni} \& {Romanova}(2009)}]{KR09}
{Kulkarni}, A.~K. \& {Romanova}, M.~M. 2009, \mnras, 398, 701

\bibitem[{{Mendigut{\'{\i}}a} {et~al.}(2011){Mendigut{\'{\i}}a}, {Eiroa},
  {Montesinos}, {Mora}, {Oudmaijer}, {Mer{\'{\i}}n}, \&
  {Meeus}}]{MENDIGUTIA+11}
{Mendigut{\'{\i}}a}, I., {Eiroa}, C., {Montesinos}, B., {Mora}, A.,
  {Oudmaijer}, R.~D., {Mer{\'{\i}}n}, B., \& {Meeus}, G. 2011, \aap, 529, A34

\bibitem[{{Mendoza}(1968)}]{MENDOZA68}
{Mendoza}, E.~E. 1968, \apj, 151, 977

\bibitem[{{Najita} {et~al.}(1996){Najita}, {Carr}, {Glassgold}, {Shu}, \&
  {Tokunaga}}]{NAJITA+96}
{Najita}, J., {Carr}, J.~S., {Glassgold}, A.~E., {Shu}, F.~H., \& {Tokunaga},
  A.~T. 1996, \apj, 462, 919

\bibitem[{{Najita} {et~al.}(2003){Najita}, {Carr}, \& {Mathieu}}]{NCM03}
{Najita}, J., {Carr}, J.~S., \& {Mathieu}, R.~D. 2003, \apj, 589, 931

\bibitem[{{Najita} {et~al.}(2007){Najita}, {Carr}, {Glassgold}, \&
  {Valenti}}]{NAJITA+06}
{Najita}, J.~R., {Carr}, J.~S., {Glassgold}, A.~E., \& {Valenti}, J.~A. 2007,
  in Protostars and Planets V, B. Reipurth, D. Jewitt, and K. Keil (eds.),
  University of Arizona Press, Tucson, 951 pp., 2007., p.507-522, ed.
  B.~{Reipurth}, D.~{Jewitt}, \& K.~{Keil}, 507--522

\bibitem[{{Najita} {et~al.}(2009){Najita}, {Doppmann}, {Carr}, {Graham}, \&
  {Eisner}}]{NAJITA+09}
{Najita}, J.~R., {Doppmann}, G.~W., {Carr}, J.~S., {Graham}, J.~R., \&
  {Eisner}, J.~A. 2009, \apj, 691, 738

\bibitem[{{Najita} {et~al.}(2000){Najita}, {Edwards}, {Basri}, \&
  {Carr}}]{NAJITA+00}
{Najita}, J.~R., {Edwards}, S., {Basri}, G., \& {Carr}, J. 2000, Protostars and
  Planets IV, 457

\bibitem[{{Rothman} {et~al.}(2005){Rothman}, {Jacquemart}, {Barbe}, {Benner},
  {Birk}, {Brown}, {Carleer}, {Chackerian}, {Chance}, {Coudert}, {Dana},
  {Devi}, {Flaud}, {Gamache}, {Goldman}, {Hartmann}, {Jucks}, {Maki}, {Mandin},
  {Massie}, {Orphal}, {Perrin}, {Rinsland}, {Smith}, {Tennyson}, {Tolchenov},
  {Toth}, {Vander Auwera}, {Varanasi}, \& {Wagner}}]{ROTHMAN+05}
{Rothman}, L.~S., {Jacquemart}, D., {Barbe}, A., {Benner}, D.~C., {Birk}, M.,
  {Brown}, L.~R., {Carleer}, M.~R., {Chackerian}, C., {Chance}, K., {Coudert},
  L.~H., {Dana}, V., {Devi}, V.~M., {Flaud}, J.~M., {Gamache}, R.~R.,
  {Goldman}, A., {Hartmann}, J.~M., {Jucks}, K.~W., {Maki}, A.~G., {Mandin},
  J.~Y., {Massie}, S.~T., {Orphal}, J., {Perrin}, A., {Rinsland}, C.~P.,
  {Smith}, M.~A.~H., {Tennyson}, J., {Tolchenov}, R.~N., {Toth}, R.~A., {Vander
  Auwera}, J., {Varanasi}, P., \& {Wagner}, G. 2005, Journal of Quantitative
  Spectroscopy and Radiative Transfer, 96, 139

\bibitem[{{Shakura} \& {Sunyaev}(1973)}]{SS73}
{Shakura}, N.~I. \& {Sunyaev}, R.~A. 1973, \aap, 24, 337

\bibitem[{{Skrutskie} {et~al.}(1996){Skrutskie}, {Meyer}, {Whalen}, \&
  {Hamilton}}]{SKRUTSKIE+96}
{Skrutskie}, M.~F., {Meyer}, M.~R., {Whalen}, D., \& {Hamilton}, C. 1996, \aj,
  112, 2168

\bibitem[{{Sun} {et~al.}(1991){Sun}, {Wu}, {Mao}, \& {Li}}]{SUN+91}
{Sun}, J., {Wu}, Y.-F., {Mao}, X.-J., \& {Li}, S.-Z. 1991, Acta Astronomica
  Sinica, 32, 134

\bibitem[{{Tannirkulam} {et~al.}(2008){Tannirkulam}, {Monnier}, {Millan-Gabet},
  {Harries}, {Pedretti}, {ten Brummelaar}, {McAlister}, {Turner}, {Sturmann},
  \& {Sturmann}}]{TANNIRKULAM+08}
{Tannirkulam}, A., {Monnier}, J.~D., {Millan-Gabet}, R., {Harries}, T.~J.,
  {Pedretti}, E., {ten Brummelaar}, T.~A., {McAlister}, H., {Turner}, N.,
  {Sturmann}, J., \& {Sturmann}, L. 2008, \apjl, 677, L51

\bibitem[{{Tatulli} {et~al.}(2008){Tatulli}, {Malbet}, {M{\'e}nard}, {Gil},
  {Testi}, {Natta}, {Kraus}, {Stee}, \& {Robbe-Dubois}}]{TATULLI+08}
{Tatulli}, E., {Malbet}, F., {M{\'e}nard}, F., {Gil}, C., {Testi}, L., {Natta},
  A., {Kraus}, S., {Stee}, P., \& {Robbe-Dubois}, S. 2008, \aap, 489, 1151

\bibitem[{{Thi} {et~al.}(2005){Thi}, {van Dalen}, {Bik}, \& {Waters}}]{THI+05}
{Thi}, W.-F., {van Dalen}, B., {Bik}, A., \& {Waters}, L.~B.~F.~M. 2005, \aap,
  430, L61

\bibitem[{{Zhu} {et~al.}(2009){Zhu}, {Hartmann}, {Gammie}, \&
  {McKinney}}]{ZHU+09}
{Zhu}, Z., {Hartmann}, L., {Gammie}, C., \& {McKinney}, J.~C. 2009, \apj, 701,
  620

\end{thebibliography}
\end{document}